# Regional Seismic Information Entropy to Detect Earthquake Activation Precursors

## Yukio Ohsawa

Department of Systems Innovation, School of Engineering, The University of Tokyo, Tokyo 113-8656, Japan; ohsawa@sys.t.u-tokyo.ac.jp; Tel.: +81-3-5841-2908

**Abstract:** A method is presented to detect earthquake precursors from time series data on earthquakes in a target region. The Regional Entropy of Seismic Information (RESI) is an index that represents the average influence of an earthquake in a target region on the diversity of clusters to which earthquake foci are distributed. Based on a simple qualitative model of the dynamics of land crust, it is hypothesized that the saturation that occurs after an increase in RESI precedes the activation of earthquakes. This hypothesis is validated by the earthquake catalog. This temporal change was found to correlate with the activation of earthquakes in Japanese regions one to two years ahead of the real activation, more reliably than the compared baseline methods.

## 1. Introduction

Methods for detecting earthquake precursors have been developed in fields relevant to earth science. The complex dynamics of Earth's land crust and its interaction with fluid have been studied, and precursory earthquake events such as nucleation, dilatancy, and colliding cascades have been modeled [1,2]. In addition, by integrating changes to wave velocity and strain, electromagnetic phenomena, and even animal behavior, the methods used for the detection of earthquake precursors have been advanced and integrated into established sciences for complex systems [3–6]. In the approach used to measure the local seismicity of each region, the appearance of seismic gaps (regions of quiescence i.e., where earthquakes are less frequent than expected based on the seismicity in the surrounding regions) may be regarded as a precursor candidate [1]. The risk of earthquakes in regions of quiescence has been shown by the Region-Time-Length (RLT) parameter, which is computed from the distribution of earthquakes based on spatiotemporal distances [7,8]. The size of a seismic gap where precursors are expected, referred to as an earthquake preparation zone, has been estimated based on deformation and tilt on the surface of the earth [9]. For comprehensive reviews of seismic precursors, see [10–12]. In these references, the debate around the prognostic value of precursors, as well as the different schools of thought, are described.

With the development of computing algorithms, purely data-driven approaches are also addressed to earthquake prediction [13–17]. For example, the eigenvectors and the corresponding eigenvalues of the $N*N$ matrix representing the pairwise co-occurrences of earthquakes in $N$ regions have been used to predict the probability of earthquake occurrences in clusters of regions [17]. Machine learning techniques used to detect the times of high change point score [18–20], based on the transition of models on latent dynamics before and after time $t$, may also have the potential to discover an essential change in land crust behavior. However, the precursors of large earthquakes have been difficult to capture using this approach because of their complex and unknown latent dynamics and extremely low frequency of occurrence. For example, the frequency of M8.0 events is $10^6$ times lower than of M4.0 events, and the precursor of the former may differ from the latter because it may be caused by larger-scale tectonic dynamics. Thus, M8.0 earthquakes cannot be predicted by learning patterns from the large data on M4.0 earthquakes.

Generally, if applied without any model of earth dynamics, a purely data-driven approach rarely works to forecast or explain "unexpected" events after they occur. Literature about unexpected earthquakes (e.g., [21–24]) show the unexpectedness of their various features, such as an unexpected timing [21], a larger magnitude than anticipated [22], or an unexpected location of focus [23,24]. As far as we specify or extend the idea to learn patterns or parameters ruling the patterns from data, it is hard to predict such events that have unexpectedness of various features and are not preceded by expectable conditions corresponding to parts of learned patterns. In general, data-based approaches in seismology have been applied to regions where earthquakes occur on a frequent basis. However, the mathematical models should be integrated with an earthquake causality model to forecast the occurrence of large earthquakes in regions where their frequency was low or in neighboring regions.

Based on the above discussion, data analysis based on models or knowledge of seismology, such as [7,8], can be a reasonable approach. For example, algorithms for clustering earthquakes on the distances in the spatiotemporal space have been shown to identify foreshocks, mainshocks, and aftershocks, and can explain their essential properties [25,26]. The value of coefficient $b$ in the Gutenberg-Richter (GR) equation has been computed from earthquake data for each year in the target region [3,27]. The value was found to decrease for a period of 10 years before an increase in the frequency of large ($M > 6.2$) earthquakes occur. However, when this knowledge is used to detect precursors, the results have been found to be unreliable for earthquakes of the smallest or the largest magnitudes. Furthermore, the changing period of 10 years means this is the time resolution we can expect in prediction. On the other hand, in statistic models of earthquake occurrences in space and time, an earthquake at each location at each time came to be modeled as the effect of previous events in the target region and surrounding areas [28–30]. In the literature on the probabilistic forecasting [28] of earthquakes, prediction within an error of 10 years was achieved for regions that experience frequent earthquakes, such as the North West and the South West Pacific Oceans. The Epidemic Type Aftershock Sequence (ETAS) also shows good performance at estimating the risk in regions where earthquakes frequently occur [29] and has been extended for use in the prediction of earthquakes of maximum magnitudes [30]. However, some earthquakes beyond the reach of these models show great exposures of energy, especially in regions where the frequency of earthquakes is low. For example, the focus of the M7.3 Kumamoto earthquake in 2016 or the focus of the M6.1 Osaka earthquake in 2018 was not captured as M7.0 or M6.0 high-risk regions by ETAS [30].

Other models used in the predictive analysis of data are found to be relevant to models in geophysics, such as theories of renormalization groups and nonlinear systems [31]. Keilis-Borok et al. modeled earthquakes as events in a nonlinear system, on which they enabled algorithmic data-based extraction of the premonitory patterns of earthquakes [5,32–35]. Their composite algorithm used a combination of patterns to predict earthquakes in various regions. For example, the CN (named after California-Nevada) algorithm was developed by a retrospective analysis of the seismicity preceding large earthquakes [35]. Here, the time of increased probability (TIP) of strong earthquakes was diagnosed using functions that represent the levels of seismic activity, the quiescence, the temporal variation of seismicity, the spatial concentration, the clustering of earthquakes, the spatial contrast of activity, and the long-range interaction of earthquakes. The interaction and the variety of earthquake activities across a wide region have been also considered in the approach of pattern informatics (PI [36,37]). Here, an earthquake is assumed to be a multi-body phenomenon ruled by latent dynamics of the lithosphere on a load plate, interacting to form a threshold system [38,39]. The value of the PI index for a region corresponding to the difference in the intensity-growth from the background regions has been found to provide forecasts of locations and the magnitude of upcoming earthquakes within an error of 10 years [40–44]. Relative Intensity [40,42,44,45], despite its computation simplicity, has been compared to (outperforming in some cases [46]) PI in the performance to detect precursors. About entropy-based analysis of earthquakes, we shall discuss in Section 2.

In this paper, the author stands on data science rather than seismology, in the sense that the focus is to find earthquake precursors from data available in an earthquake catalog. Nevertheless, as discussed above, we should take a simple model of earthquake dynamics into account to forecast the unexpected activation of earthquakes. Here, we borrow the idea of a cluster-based analysis of data from an earthquake catalog [47], where the co-occurrence of earthquakes has been used to extract not only the clusters of active faults rupturing within a small time frame but also the relationships among multiple clusters. In [47], several regions with low earthquake frequency were highlighted as "near-future risk" and coincided with the locations of real events that occurred later. This method is based on a simple model of earthquakes, assuming that such a region may be stressed by the movements of multiple clusters of active faults. However, information regarding temporal changes in the data is eliminated when the method relies on the co-occurrences as a statistic quantity. In this paper, to detect a short-term (preceding 1 year or 2 year) precursor of the activation of earthquakes in a target region, a quantitative index called the Regional Entropy of Seismic Information (RESI) is proposed. RESI is based on a simple hypothetical model of land crust dynamics proposed in the next section. It extends the idea of inter-cluster interaction by introducing the temporal transition of the diversity of clusters. The performance of RESI concerning the detection of earthquake precursors is evaluated on the data in an earthquake catalog.

## 2. Restructuring of Earthquake Foci Clusters: A Simple Model of the Precursory Process

Let us introduce a simple model of land crust dynamics to explain the precursory process involved in the activation of earthquakes as a basis for the data analysis. The model assumes the transitions from state (a) or from (b) to (e) (expressed {(a) or (b)} to (e) hereafter) illustrated in Figure 1 across the entire $S^U$ geographical region covered by the target data. It is composed of the two phases described below. In this simple model, we investigate the dynamic restructuring of clusters of earthquake foci (approximated by "quaking meshes" later), including the separation/combination of clusters and the activation/deactivation of earthquakes, based on the data about the time and the location of each earthquake.

**Phase 1**: The diversity of clusters to which the foci of earthquakes distribute increases from {(a) or (b)} to state (c) in Figure 1. Here, a cluster comes to be separated to create a seismic gap (as shown in Figure 1c) because of the local disappearance of earthquakes in the central part of (a), or the appearance of new cluster(s) beside the existing one such as the shift from (b) to (c).

**Phase 2**: Earthquakes converge to a smaller number of clusters (from Figure 1c–e) possibly via state (d). Here, the clusters of foci are combined (as shown in (e)) if earthquakes occur in the seismic gap in (c). In this step, the tentative seismic gap shown in (c) becomes the preparation zone for earthquakes in the transition to (e). Before reaching state (e), earthquakes may occur in the seismic gap and the state may move closer back to (c) or forward to (e) via (d) during the transition period. If (e) is reached, the clusters linked via the bridge are combined.

Among the hypothetical mechanisms available to explain the appearance of seismic gaps (i.e., the precursory quiescence as surveyed or modeled in [10–12,48]), the above two-phase model may be interpreted on the relevance to the locked fault model. According to this model, a segment of the creeping fault is partially locked. As a result, the creep rate and the seismicity rate are reduced in the fault zone rather than in the surrounding land crust. In Figure 1d, in the seismic gap, a small activation may restart before the mainshock. Observations shown in the literature may be interpreted as events on the bridge in (d), such as the appearance of some of the earthquake swarms in a seismic gap [49]. However, the bridge in (d) may not always occur and may be a transient phenomenon; that is why the direct arrow from (c) to (e) is shown along with the path via (d) in Figure 1. However, we do not exclude other models such as the slip softening model where the precursory creep near the horizontal subsequent mainshock fault plain lowers the ambient stress in the crustal volume above the fault plane. According to this model, the quiescence is distributed in the crustal volume. This quiescent crustal volume may be associated with the area that turns into a quiescence during the transition from (a) to (c) in Figure 1, which is wider than the bridge in (d). Thus, the author does not choose a specific causal model for the birth of a seismic gap in (c). Instead, the author just assumes that the region may share a latent root cause with the ambient regions where earthquakes are active. If a seismic gap appears as in Figure 1c or earthquakes begin to occur in the seismic gap as shown in (d), it is regarded as a precursor of the mainshock.

Thus, let us summarize the precursory pattern that should be extracted. First, a seismic gap appears in Figure 1c due to the shift to quiescence in the central region of (a) or the growth of clusters from the state of (b). This appearance of a seismic gap surrounded by the clusters of active earthquakes as in Figure 1c comes to be released as in (d) or (e) until a new cluster emerges in (e). This process of clusters' restructuring, capturing the dynamics in Figure 1, can be represented by information entropy [50] $H(S^U, t)$ in Equation (1) for the entire region $S^U$ at time $t$. $p(C_i|S^U, t)$ denotes the conditional probability that an earthquake occurred in cluster $C_i$ under the condition that its location was in $S^U$ at time $t$. $C_i$ is the $i$-th cluster of earthquake foci in $S^U$.

$$H(S^U, t) = \sum_i p(C_i|S^U, t) \log p(C_i|S^U, t) \tag{1}$$

In this specification of entropy, the $i$-th microstate is interpreted as the $i$-th cluster $C_i$ of foci in region $S^U$ to which the focus of each earthquake may belong. The definition of a cluster will be given later in the proposed algorithm and can be understood as a group of closely located foci of earthquakes in a period (e.g., 1 month or 1 year).

The value of $H(S^U)$ increases in the transition from the state of Figure 1{(a) or (b)} to (c). If the density on the bridge is unstable, a perturbation of entropy may be found. This is because $H(S^U)$ increases in the transition from state (c) to (d) if the foci on the bridge do not cluster together in (d) and decrease in the opposite transition. When reaching (e), or if the epicenters on the "bridge" of (d) increase to a sufficiently large density, $H(S^U)$ decreases because clusters come to be united. It is possible that some clusters in (c) disappear, which manifests as a decreasing pattern in $H(S^U)$. Such a case is regarded as a transition represented by the dotted arrow from (c) to (b), followed by a return to (c) if the mainshock follows. Thus, state (c) and (d) are assumed to imply the existence of a precursor of the activation of earthquakes in all the transition paths considered here.

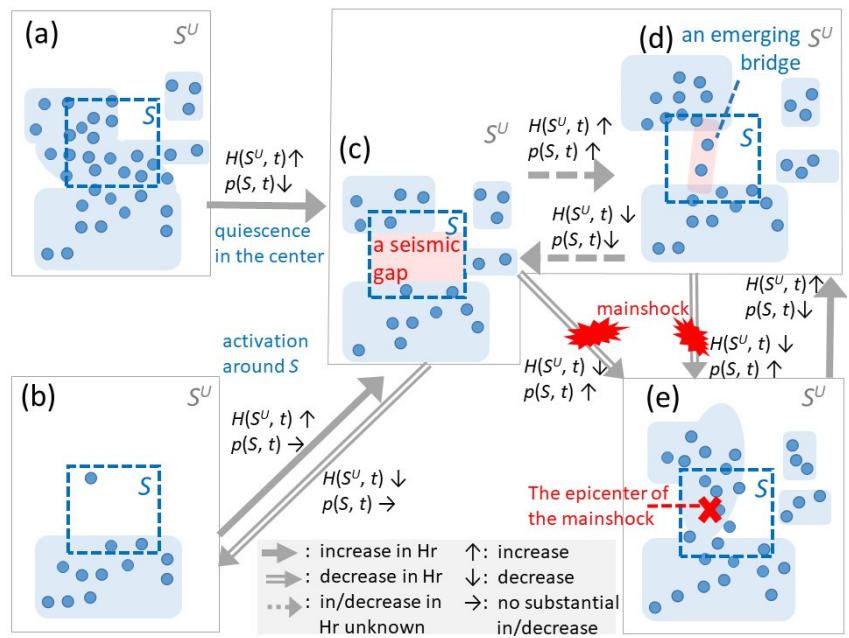

**Figure 1**. Transitions of earthquake activation. In each state from (**a**) through (**e**), the appearance of seismic gaps and bridges are regarded as precursor candidates for earthquake activation. Hr($S$, $t$) and $p(S, t)$ here are referred to in Section 3 where RESI is defined based on $H(S^U)$.

Although the above model may be too simple to capture the dynamics of actual earthquakes, it is expected that its simplicity absorbs the differences between the dynamics of various types of earthquakes. For example, deeper earthquakes on the dipping interface of subduction plates and shallower earthquakes on active faults in the inland are ruled by different dynamics. If we apply fine models to analyze the activities of land crust to reflect such differences, we should build various models and a hybrid model if earthquakes occur from a mixture of dynamics.

*A Note about Entropy*

Please note that the concept of entropy introduced in this paper is not the same as that used in statistical physics [51] or its application to understanding earthquakes [52]. The entropy concept discussed here is that discussed in information science. In statistical physics, entropy has been defined as being related to the numbers of microscopic configurations of a thermodynamic system, specified by macroscopic variables. For example, in thermodynamics, entropy can be specified by a physical parameter of the system as energy divided by temperature. In statistical mechanics, entropy is a measure of the number of ways in which a system can be arranged, quantified by the average logarithm of the number of possible microscopic configurations of the particles in the target system. It is observed as a macroscopic disorder. On the other hand, in information science, entropy is defined as the sheer amount of information needed to specify the full set of microstates of the system, not specified by a macroscopic variable. In this paper, entropy refers to the definition in information science and regards the clusters defined below as microstates. Thus, entropy here is introduced to quantify the diversity of clusters of "quaking meshes", land meshes in the target region that include foci. Clusters of quaking meshes are introduced as an approximation of earthquake foci clusters.

Let us position this method within the context of how information entropy has historically been applied. First, the proposed method is intended to compute entropy based on clusters to quantify the diversity of earthquake foci, not to use entropy as a criterion for clustering, as is the case in some literature [53]. This point is similar to the use of entropy in marketing, where entropy has been used as an index for the diversity of interests and products [54–56]. In digital images, the temporal change of entropy for each part of a given image has been used to detect contours and their movements [57]. Furthermore, entropy in traffic and events in computer networks have been demonstrated to provide a scalable technique to detect unexpected behaviors and abrupt changes [58,59]. In this history, the contribution of this paper is to obtain entropy on the development of a method for clustering foci and to consider the average influence of every single earthquake in the target region to the entropy of the entire region $S^U$ as a quantitative measure of earthquake precursors.

In recent analyses of earthquakes, the complexity measure associated with the entropy change of seismicity under time reversal has been found to occur before the occurrence of a major earthquake [60,61]. Here, each earthquake in the time series is addressed as a microstate. In the context of Tsallis Entropy [62–67], the entropic index $q$ that expresses the degree of non-extensivity of the system has been shown to represent the magnitude-frequency distribution, the spatiotemporal properties of earthquake swarms, asperities, and the existence of regional hydrothermal features. Natural time analysis revealed that the Tsallis formulation achieves a satisfactory description of real seismic data for Japan when the index is supplemented by long-range temporal correlations [63]. The temporal change in $q$ has also been shown to grow gradually and then exhibit an abrupt increase upon the occurrence of a large earthquake [61]. A specification of entropy based on the distribution of epicenters on the land have also been proposed, but here we do not go into details because its aim was to measure and reduce the disorder of earthquake distributions [68]. On the other hand, in this paper, clusters of earthquake foci are introduced as microstates when computing entropy. In addition, we focus on the saturation of RESI, derived from $H(S, t)$ in the next section, at the maximum value. The author does not say this is the best method, but that the cluster-based entropy can give a suggestion about the risk at a given time, in addition to other methods. Therefore, the author is planning to continue to discuss the relations of RESI and other specifications of entropy from the aspect of earthquake dynamics to find a setting for either a meaningful comparison or a constructive combination with other methods.

## 3. Regional Entropy of Seismic Information

### 3.1. Definition

Let us model the restructuring dynamics of the cluster (as shown in (a) through (e) in Figure 1) by developing the RESI $Hr(S, t)$ in Equation (2).

$$Hr(S, t) = H(S, t) - \log p(S, t) \tag{2}$$

Here, $H(S, t)$ is given by $\sum_{C \subset Msh(S)} p(C \,|\, S, t) \log p(C \,|\, S, t)$, which is obtained by replacing $S^U$ and the foci in Equation (1) with a region $S$ in $S^U$ and the meshes in $Msh(S)$, where $Msh(S)$ is the set of "quaking meshes", that are meshes of land including earthquake foci in $S$. Here, foci locations are approximated by quaking meshes. In addition, we approximate the 3D location of earthquake foci in 2D by ignoring the depth. Although depth consideration is expected to improve the performance of the presented method, we approximate its value in this manner to a create a fair comparison with the baseline methods that essentially use 2D information in the data. That is, quaking meshes are the meshes including epicenters.

$Hr(S, t)$ in Equation (2) can be rewritten as $\sum_{C \subset Msh(S)} p(C, t) \log p(C, t)$ divided by $p(S, t)$ based on the assumption that $p(C, t)$ is equal to $p(C \,|\, S, t) \, p(S, t)$. This means $C$ belongs to $S$, and $\sum_{C \subset Msh(S)} p(C \,|\, S, t)$ is equal to one, which is similar to the assumption made in Appendix A. Here, for the entire given map $S^U$, $p(C)$ means $p(C \,|\, S^U)$. In addition, $H(S^U, t)$ in Equation (1) is the sum of $\sum_{C \subset Msh(S)} p(C, t) \log p(C, t)$ for all $S$ in $S^U$. Thus, $Hr(S, t)$ represents the contribution of clusters in $S$ to the entropy of $S^U$, divided by the rate of earthquakes in region $S$ among all the clusters in $S^U$. In this sense, intuitively, $Hr(S, t)$ indicates the average contribution per earthquake in region $S$ to the diversity of clusters of foci in $S^U$.

As in shown in Figure 1, $H(S^U, t)$ substantially increases and $p(S, t)$ decreases or stays without a substantial change in the transition from {(a), (b), or (e)} to (c). In contrast, $H(S^U, t)$ substantially decreases and $p(S, t)$ increases if some earthquakes in the seismic gap connect clusters in (c), urging forward to (e). The former and the latter cases result, respectively, in a substantial increase and decrease in $Hr(S, t)$. Therefore, if $Hr(S, t)$ is converging to its maximum value, it can be expected that the transition of clusters in $S^U$ to (c) is in the final stage. This convergence can be observed as a saturation of the increase in $Hr(S, t)$. If a perturbation of $Hr(S, t)$ is observed after this saturation, this implies the uncertainty of the in/decrease in $Hr(S, t)$ (because $H(S^U, t)$ and $p(S, t)$ change in the same way) depicted in the dotted arrows between (c) and (d) in Figure 1 or the unstable changing in $Hr(S, t)$ between (d) and (e). Both causes of the perturbation of $Hr(S, t)$ occur due to the perturbation of the earthquake density on the bridge in state (d), that causes the transition forward closer to state (e) or back closer to (c). Thus, we expect to detect precursors of earthquake activation in region $S$ based on the increase, the saturation, and the perturbation of $Hr(S, t)$ in Equation (2).

In addition, $Hr(S, t)$ for region $S$ can be computed as the average of $Hr(S_i, t)$ for all regions $S_i$ in $S$, as shown in Appendix A. This makes it possible to conveniently compute the value of RESI for region $S^*$, which represents a union of subregions, such as $S_i$, by a linear computation.

### 3.2. The Algorithm Used to Obtain Alarms of Precursor Candidates on RESI

The value of RESI is obtained in two steps. In Step 1, the clusters of quaking meshes (square areas of 0.1° of latitude and longitude including the foci of earthquakes occurred over a cutoff frequency) in the target region $S$ are generated, and in Step 2 the value of RESI (i.e., $Hr(S, t)$) is computed. Here, let us divide the target region $S$ into meshes of size $(\Delta x, \Delta y)$ that are to be clustered in Step 1, where $\Delta x$ and $\Delta y$ are the widths in latitude and longitude, respectively. $x_L$ and $y_L$ represent the widths in latitude and longitude of region $S$, respectively, which are substantially wider than the meshes of width $\Delta x$ and $\Delta y$. $[t, t + t_L]$ is the time range for which $Hr(S)$ is computed by setting $t_L$ to 1 year or 1 month. In Step 1–2 (the second sub-step of Step 1) below, $Msh(S)$ is given as the set of meshes where a larger number of earthquakes than a given $\theta_m$ of magnitude $M_\theta$ or larger occurred in the period $[t, t + t_L]$. The cutoff magnitude $M_\theta$ is set to 2.0 for the reasons mentioned in **Data and Their Availability** referring to Appendix B, accepting the possibility that they may be aftershocks of previous events. $\theta_m$ is set to one to avoid taking meshes with only one earthquake. $p(X, t)$ for region $X$ or cluster $X$ is computed as quakes($X$, $t$)/quakes($S^U$, $t$), where quakes($X$, $t$) is the number of earthquakes in $X$ of magnitude $M_\theta$ or greater in period $[t, t + t_L]$. Let us hereafter represent a point on the Earth by ($x$: latitude, $y$: longitude) skipping the SI unit (°), and a rectangular region of four vertices ($x,y$), ($x', y$), ($x, y'$), ($x', y'$), by ($x,y$)-($x', y'$).

The clustering function **make_clusters**($S$, $t$) called in Step 1–3 runs as follows (see Appendix C, where [69–72], Figures A2 and A3, and Appendix B are referred to). Each cluster $C_{x0y0}$ grows as a subset of Msh($S$) (given in Step 1–2) from a seed mesh $m_{x0y0}$ selected randomly from quaking meshes belonging to no cluster, by following and absorbing quaking meshes in the neighbors not belonging yet to any cluster. If there are the neighbors of mesh $m_{xy}$ in $C_{x0y0}$, that are members of Msh($S$) but do not belong to any cluster generated so far, adding those neighbors to $C_{x0y0}$ is called to extend $m_{xy}$. In addition, the meshes already extended are called "Extended." This cycle of seeding and growing clusters via extending meshes in them is iterated until all meshes in Msh($S$) are covered by the generated clusters. Here, the representation of a cluster by $C_{x0y0}$ means that it is represented by its starting seed mesh $m_{x0y0}$. In addition, surround($m_{xy}$) is the set of eight meshes surrounding $m_{xy}$: $\{m_{x-\Delta xy+\Delta y}, m_{xy+\Delta y}, m_{x+\Delta xy+\Delta y}, m_{x-\Delta xy}, m_{x+\Delta xy}, m_{x-\Delta xy-\Delta y}, m_{xy-\Delta y}, m_{x+\Delta xy+\Delta y}\}$. On the clusters obtained in Step 1, RESI is computed as Hr($S$, $t$) in Equation (2), on which alarms are obtained as Hr$_{sat}$($S$, $t$) below by excluding the "else" condition from times of positive Hr($S$, $t$), as in Equation (3).

**(Step 1: obtain clusters of quaking meshes in the target region $S$ for the period $[t, t + t_L]$)**

1–1) Divide the target region $S$, ($x_0$, $y_0$) − ($x_0+x_L$, $y_0+y_L$), into meshes of a given size ($\Delta x$, $\Delta y$).

1–2) For the period $[t, t + t_L]$, take Msh($S$), the set of quaking meshes in $S$.

1–3) Do **make_clusters**($S$, $t$) below.

**(Step 2: obtain RESI)** Obtain RESI as Hr ($S$, $t$) using Equation (2). $C$ is a cluster of quaking meshes in $S$, and $p(C\,|\,S, t)$ denotes the conditional probability that an earthquake occurred in $C$ under the condition that its location was in $S$ for the period $[t, t + t_L]$. $p(C\,|\,S, t)$ is the division of the number of earthquakes in $C$ by the number in region $S$ in $[t, t + t_L]$.

**(Step 3: obtain the alarms of candidates of precursors)** The time when the value of Hr$_{sat}$($S$, $t$) defined below is larger than zero, i.e., when Hr($S$, $t$) saturates at the highest value range of the last period of length $T$, is taken as an alarm of the precursor of earthquake activation.

**make_clusters**($S$, $t$)
    Extended = {} (i.e., empty set);
    while Msh($S$) ≠ {}:
        $m_{x0y0}$: = a randomly selected member of Msh($S$); $C_{x0y0}$ = {$m_{x0y0}$}
        For each mesh $m_{xy}$ in $C_{x0y0}$/Extended: #*each mesh not extended yet gets extended below*
    while Msh($S$) ∩ surround($m_{xy}$) ≠ {}
    add_to_cluster = Msh($S$) ∩ surround($m_{xy}$)
        $C_{x0y0}$ = $C_{x0y0}$ ∪ add_to_cluster
        Msh($S$) = Msh($S$)/add_to_cluster
        Extended = Extended + $m_{xy}$
return
**Computing Hr$_{sat}$ ($S$, $t$) as the alarm:**
    if    rank $_{\tau \text{ in } [t\text{-min}(T, t\text{-}t0), t]}$ Hr$_{avr}$($S$, $\tau = t$) ≤ $\gamma$ min($T$, $t$-$t_0$) and
        (stdev $_{\tau \text{ in } [t\text{-}dt, t]}$ Hr($S$, $\tau$) < $\theta_{std}$ or stdev $_{\tau \text{ in } [t\text{-}dt/2, t]}$ Hr($S$, $\tau$) > 2 stdev $_{\tau \text{ in } [t\text{-}dt, t\text{-}dt/2]}$ Hr($S$, $\tau$))

        Hr$_{sat}$($S$, $t$) = Hr($S$, $t$)

    else        Hr$_{sat}$($S$, $t$) = 0

$$(3)$$

Here, Hr$_{avr}$($S$, $t$) is the average of Hr($S$, $t$) for the preceding 6 months $[t − 5, t]$. The first and the second lines represent, respectively, a top-ranked value of Hr$_{avr}$($S$, $t$) at time $t$ in the last min($T$, $t − t_0$) years where $T$ is given and $t_0$ is the starting time of the target data (Jan 1983 for the results below), and the reduction of temporal variation of Hr($S$, $t$) due to its saturation in (c) of Figure 1 **or** the perturbation of Hr($S$, $t$) during the transition period via (d) in Figure 1. $T$ is set to 28 year in the experiments (except for Figure A8) for the reason discussed in Section 5. $dt$ is set to 1 year. In the case where $t_L$ is set to 1 year, Hr$_{avr}$($S$, $t$) is replaced with Hr($S$, $t$), and stdev $_{\tau\,in\,[t-dt,\,t]}$ (Hr($S$, $\tau$)) is replaced with |Hr($S$, $t$) − Hr($S$, $t$ − 1 y)| because only two data points are in [$t$ − 1 y, $t$]. $\theta_{std}$ is set to 0.5, $\gamma$ to 0.1 in the experiments below.

## 4. Results

### 4.1. The Data on Earthquakes

We used data on the $0.61 \times 10^6$ earthquakes in Japan from 1983–2017. The cutoff $M_0$ was set to 2.0 (see **Data and Their Availability**, Figure A4, and Table A1 about these choices). Regions dealt with in this paper are in the entire region of (25,125)–(49,149) as shown in Figure 2. This region includes a major portion of the islands of Japan, where the seismographs used to collect the data are located. This square was divided into 36 cells, and each cell corresponds to a region of 4° in latitude and longitude. For example, the bottom left-cell is (25,125)–(29,129).

According to the author's analysis from the data on estimated errors (see Figure A4 in the appendix), the average for all the years in 1983–2017 of the yearly average error of epicenters is $(1.19,1.59) \times 10^{-2°}$ based on the Japan Meteorological Agency (JMA) catalog. Furthermore, the standard deviation of the average error is $(1.13,1.71) \times 10^{-2°}$ for (latitude X, longitude Y). The mesh size of the 0.1° square is the finest resolution in the range where the errors of the epicenters can be absorbed because this is larger than the average error by more than five times of the standard deviation. The epicenter location error has been suspected to be on the order of 10 km (close to 0.1° in latitude and longitude) in the literature [73], but the JMA data shown here depicts even less error.

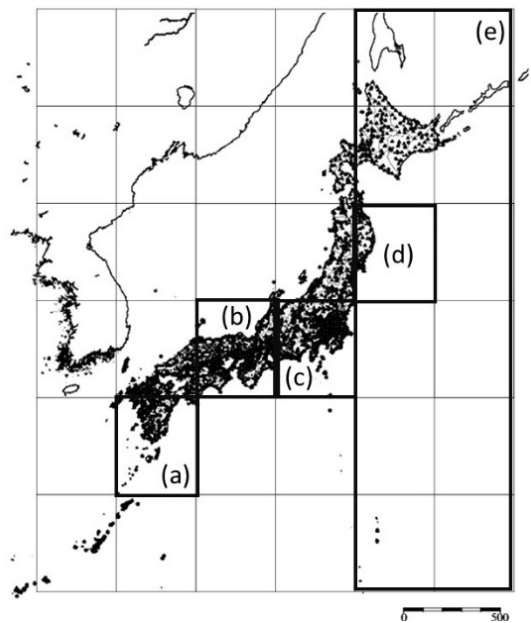

**Figure 2.** Division of (25,125)–(49,149) into 36 regions of 4° in latitude and longitude. The proposed method is evaluated for regions (**a**) through (**e**) in this figure. Each region of 4° square is called a cell. The dimensions of the cells have been simply chosen based on latitude and longitude, excluding the consideration of geological or geographical features, to purely focus on the effect of the present RESI-based method. The small dots in the land area represent the locations of seismographs. The SI unit for degree (°) is skipped in the representation (25,125) in this paper. Source of the background map: Japan Meteorological Agency website (http://www.data.jma.go.jp/svd/eqev/data/intens-st/). The modification and the use of this map are licensed as in http://www.jma.go.jp/jma/en/copyright.html.

## 4.2. The Time Series of Clusters and RESI

In this section, let us show several results of computing RESI. $(x_L, y_L)$ is set to $(4°, 8°)$ to address regions (a) through (d) and $(24°, 8°)$ to address regions (e) that includes (d) in Figure 2. $t_L$ is set to 1 year or 1 month in the experiments, and $(\Delta x, \Delta y)$ is set to $(0.1°, 0.1°)$. In Step 1, the clusters of quaking meshes are obtained as shown in Figure 3. In Step 2, $\mathrm{Hr}(S, t)$ is obtained for each panel of (a) through (e) in Figures 4 and 5, which correspond to each region of (a) through (e) in Figure 2. Here, RESI is found to increase and then saturate a few years before the activity($S, t$) in Equation (4) increases toward peak values. Equation (4) represents the magnitude that corresponds the total energy of all earthquakes, $N_{t,s}$ represents the number of earthquakes, and $M_{t,s,k}$ represents the magnitude of the $k$-th earthquake in region $S$ in the time unit $[t, t + t_L]$. 31.62 is the logarithmic bottom for the JMA magnitude.

$$\text{activity}(S, t) = \log_{31.62} \sum_{k=1}^{N_{t,S}} 31.62^{M_{t,s,k}}. \tag{4}$$

In Figure 3a–e, we observe the clusters obtained in Step 1 and the precursor alarms obtained as $\mathrm{Hr_{sat}}$ above for the five regions in the thick frames of Figure 2a–e. Then, we find the correlation of $\mathrm{Hr}(S, t)$, $\mathrm{Hr_{sat}}(S, t)$, and activity($S, t$) by setting $t_L$ to 1 year in Figure 4 and 1 month in Figure 5, where panels (a) through (e) correspond to (a) through (e) in Figures 2 and 3 as follows. The mainshocks are marked by a red cross in Figure 3. The alarm signals in Figures 4 and 5 are the times where the blue lines, representing $\mathrm{Hr_{sat}}$, take positive (non-zero) values.

(a)   In the change from 1993–1995 in the region corresponding to the cell of Figure 2a, new clusters emerge until 1995 as shown in Figure 3a. This corresponds to the increase in $\mathrm{Hr}(S, t)$ depicted in Figures 4a and 5a, which saturated at (a-2) in 1994 before the M6.9 earthquake occurred in October 1996 at (a-1). Thus, the target region itself came to form the state of Figure 1c in 1994, followed by (d) in 1995 and (e) in 1996.

(b)   Corresponding to the region of Figure 2b, we see an increase in clusters until 2000 as in Figure 3b. By 2000, RESI is saturated as shown in Figures 4b and 5b. The increased clusters are located far from the focus of the M7.3 earthquake in October 2000, that corresponds to (b-1) in Figures 4 and 5. However, as shown in Figure A5 in the appendix, the area of intense quaking caused by this earthquake ranged in the areas of these clusters. The alarm of (b-2), obtained as the time of non-zero $\mathrm{Hr_{sat}}(S, t)$, preceded (b-1) by one year in both Figures 4 and 5.

(c)   In Figure 3c, we observe an increase in clusters until 1999, which is when saturation occurs at (c-2) of RESI in Figures 4c and 5c. Then, the M6.1 earthquake occurred in August 2000, which reduced RESI substantially, as shown in (c-1) in both Figures 4c and 5c.

(d)   (and (e)) In Figure 3e, corresponding to the wide region Figure 2 e including region (d), we find at least two clusters in the ocean came to be united in 2011. One cluster existed since before 2006, and the other cluster occurred where earthquakes increased since 2006. The main M9.0 shock occurred on 11 March 2011 at the red cross between these clusters, then the clusters united. However, in Figure 4d, we find no precursory saturation for the M9.0 earthquake at (d-1). On the other hand, we find that RESI saturated at (e-2) in Figure 4e, where (e-1) represents the M9.0 earthquake. In Figure 5, we also find the precursor (e-2) for (e-1), but (d-2) is not so close to (d-1) as (e-2) is to (e-1). These results will be discussed in the Discussion section.

Thus, between the two time-scale settings (1 year and 1 month) of $t_L$, we basically find similar correlations between RESI and the activation of earthquakes. However, as shown in (d) versus (e) in Figures 4 and 5, we should set a suitable scale $(x_L, y_L)$ for the land. We shall return to this point in Section 5.

Furthermore, the reliability of RESI as a method for precursor detection should be evaluated based on false negative and positive. Here, a false negative case means a time $t$ when an alarm (a positive value of $\mathrm{Hr_{sat}}(S, t)$) is not found although a peak of activity (here defined as an activity which is in the top $k$ in the curve and is the largest in the period of $[t - 2$ year, $t]$, where $k$ is the number of obtained alarms) exists within the period of $\Delta t$ after $t$. A false positive means a time $t$ when an alarm is found but no peak of activity exists within $\Delta t$ after $t$. For example, in the visualized curves in Figure 4, false negative cases are the periods of 2 year before 2009 in (a), none in (b), 1988–1989 and 2 year before 2009 and before 2014 in (c), none in (d), and none in (e) if the allowed gap ($\Delta t$) is set to 2 year. For the equal value of $\Delta t$, false positives in 2011 are (a), the period from 1990 till 1992 in (b), 1991 till 1994 in (c), none in (d), and 2007 in (e). As far as we find these errors, the evaluation from this aspect should be discussed. In Section 5, we discuss the performance of RESI introducing two functions that correspond to the lowness of false positive and false negative error rates, with setting $t_L$ to 1 month for taking a larger number of sample times and a finer time resolution.

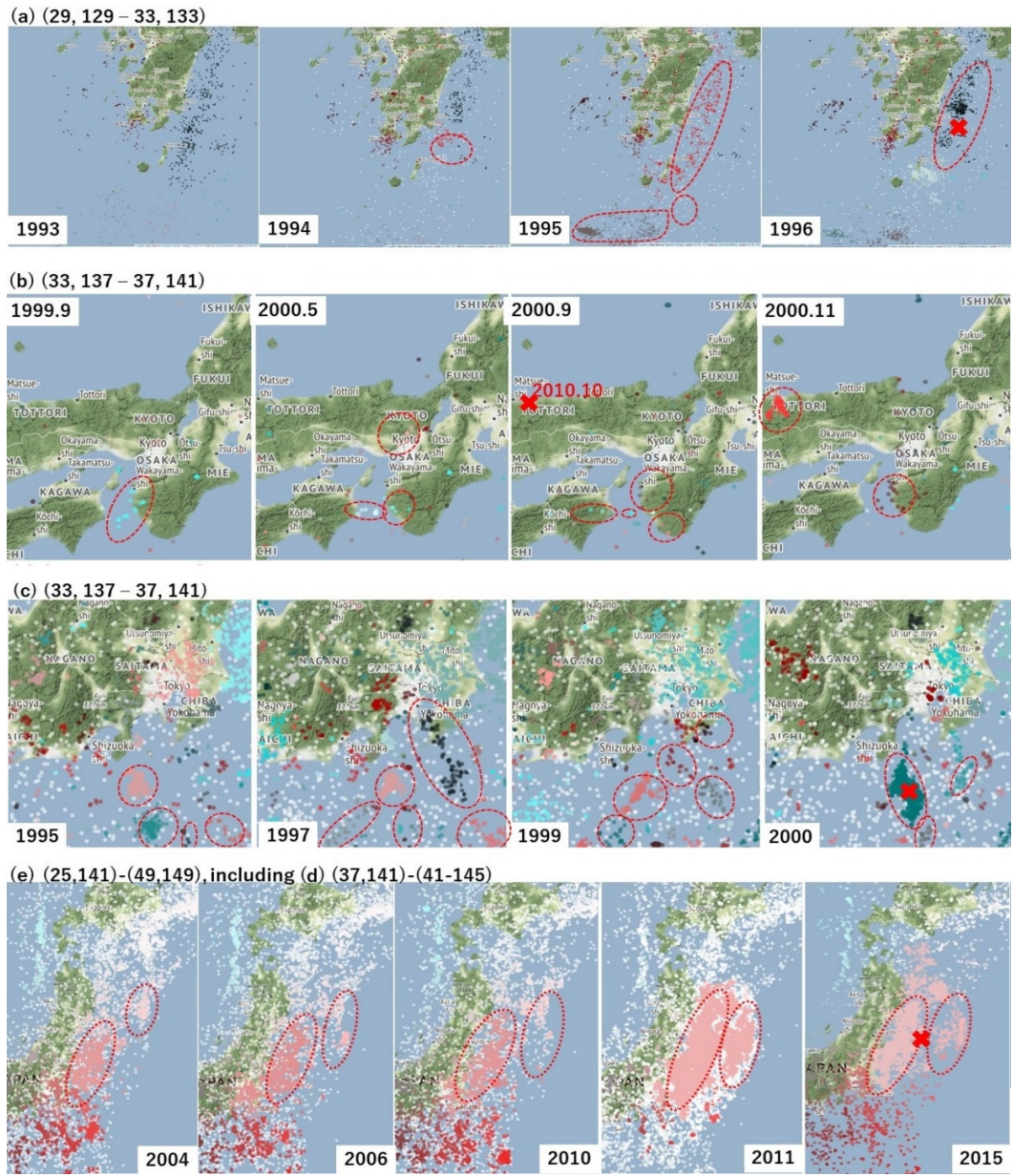

**Figure 3.** Running examples of Step (1): clusters obtained. The colors in (**a**) through (**e**) show the clusters of quaking meshes obtained in Step 1 for the regions of Figure 2a–e. The red crosses show the epicenters of large earthquakes corresponding to the peaks of activity in Figures 4 and 5. The dotted ellipses show the clusters referred to in the text. The maps have been created using Folium copyrighted since 2013 by Rob Story, licensed under the MIT License (https://github.com/python-visualization/folium/blob/master/LICENSE.txt).

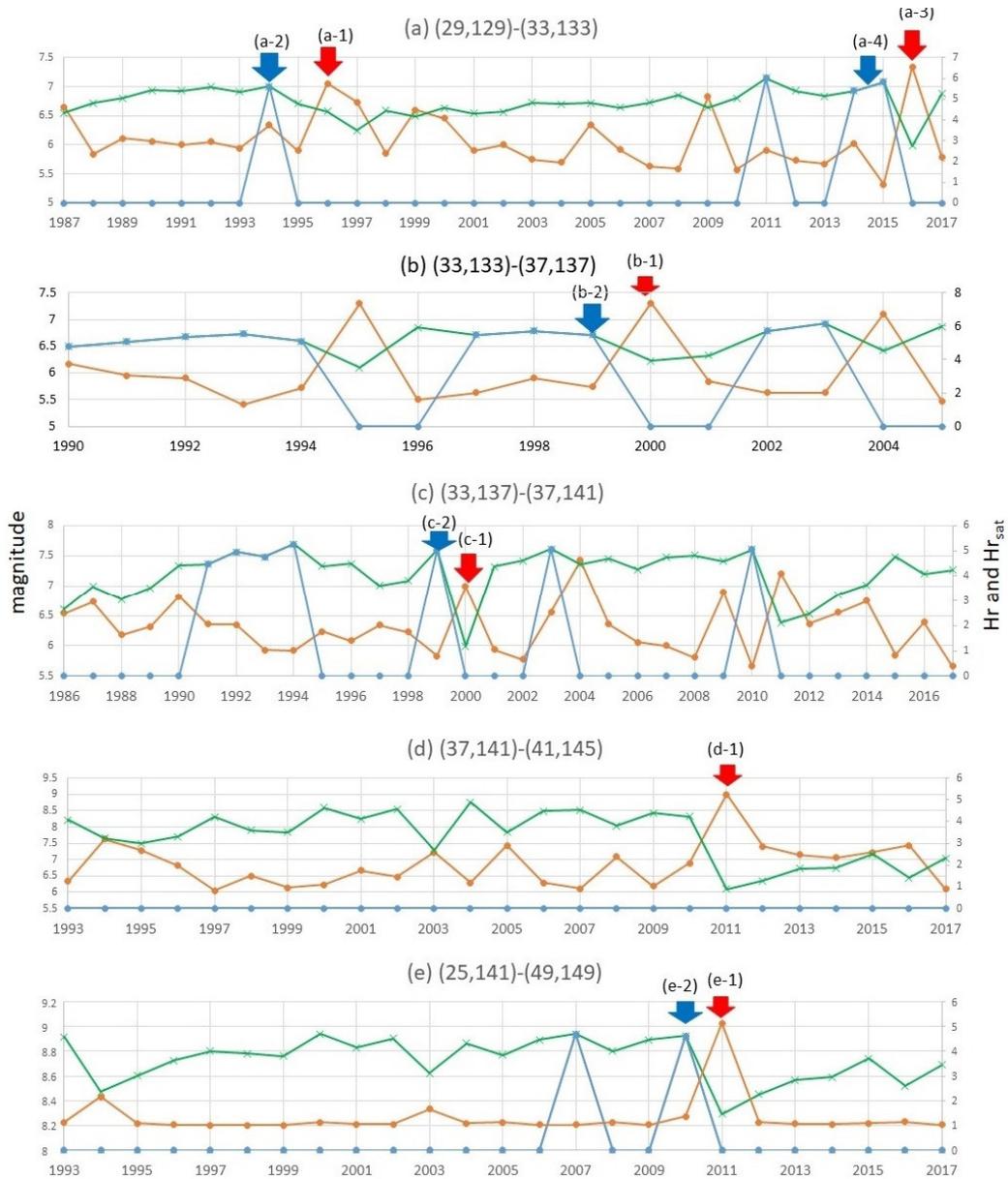

**Figure 4.** The curves of alarms on RESI and earthquake activation for $t_L$ set to 1 year: RESI (Hr($S$, $t$), green line), the precursor alarm obtained as Hr$_{sat}$ (blue line), and the earthquake activity (orange) of each region. Panels (**a**) through (**e**) correspond to the five regions in Figure 2. The times of (**a-3**) and (**a-4**) are referred to in Section 5 later.

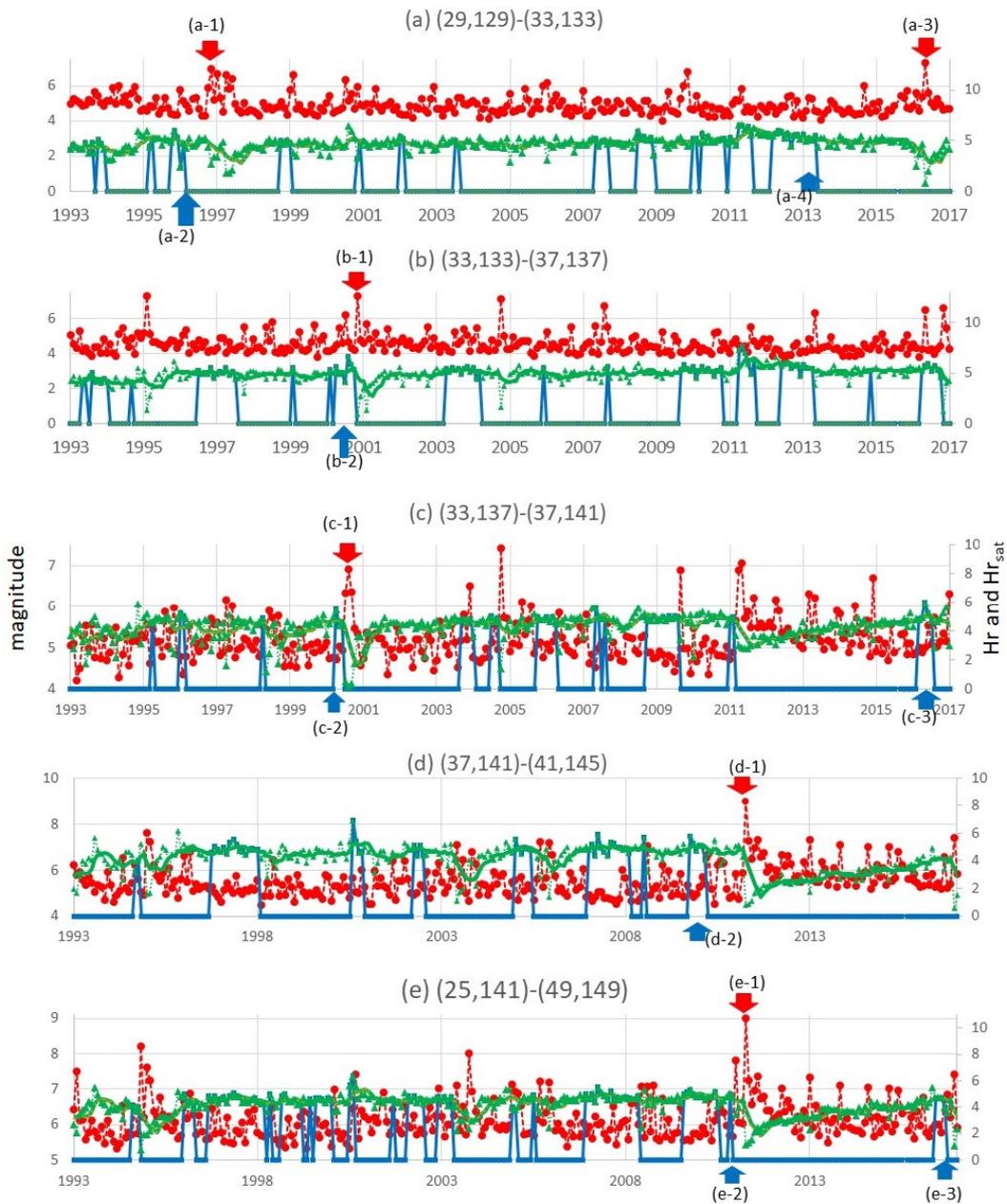

**Figure 5.** The curves of alarms on RESI and earthquake activation for $t_L$ set to 1 month: setting other conditions similar to Figure 4. (**a-3**), (**a-4**), (**c-3**), and (**e-3**) are referred to in Section 5.

## 5. Discussions

### 5.1. The Curves of Alarms on RESI and Earthquake Activities

The curves in Figures 4 and 5 suggest that the peaks of RESI tend to precede the peaks of earthquake activity. As previously mentioned, we found that RESI became saturated at (e-2) in Figure 4e, where (e-1) represents the M9.0 earthquake. However, we do not find such a precursor in Figure 4d. A similar tendency was found in Figure 5. Considering that the wide frame (e) in Figure 2 covers not only the red cross in Figure 3e but also the epicenters of all the earthquakes larger than M7.0 within an hour after the M9.0 mainshock [74] (which region (d) did not cover), it can be assumed that the mainshock occurred from the dynamics of an area wider than a cell in Figure 2. This example shows that we should choose a suitable land scale for evaluating risks on RESI.

Some failures are found in the results. For example, (a-3) does not appear within a few years after (a-4) in Figure 5 as it does in Figure 4. This difference occurs because of the differences for setting $t_L$ to 1 month and to 1 year. Therefore, we should say the selection of $t_L$ can affect the results. The change in Hr($S$, $t$) is less stable for shorter $t_L$ because the number of events is smaller in the period [$t$, $t_L$]. This is compensated for by taking the average as Hr$_{avr}$($S$, $t$). On the other hand, in Figure 5, we find a decrease in Hr for the 3 year period after (a-4) before a large earthquake occurs at (a-3). This is an exceptional phenomenon in that Hr normally tends to increase during long time scales, as in Figure A7. Such a phenomenon has not been considered yet in Figure 1.

*5.2. Comparison with Baselines*

Here, let us compare the utility of Hr$_{sat}$($S$, $t$) with baseline alarming methods by computing their correlation (embracing delay or precedence within a given range of time gap) with activity($S$, $t$) for each region $S$ that is a cell in Figure 2. In this section, to collect a larger number of sample times and to evaluate for the precedence or the delay in a finer resolution than 1 year, we set $t_L$ to 1 month. In preparation, let us define meta-function high_$f$ for an alarming function $f$ for each region $S$ as follows.

For Hr as $f$,

$$\text{if Hr}_{sat}(S, t) > 0: \qquad \text{high\_Hr}(S, t) = 1$$
$$\text{else:} \qquad \text{high\_Hr}(S, t) = 0 \tag{5}$$

For another function (baseline to be compared with RESI) that is to be PI or relative intensity (RI) summarized in Appendix D on the references [36–46], high_$f$($S$, $t$) is set to 1 for $ts$ of the largest $n$ values of $f$($S$, $t$), where $n$ is equal to $mT_f / T_{Hr}$ for $m$. For other $ts$, high_$f$($S$, $t$) is set to 0. Here, $m$ is the number of all $t$ where Hr$_{sat}$($S$, $t$) > 0. This choice of the largest $n$ values intends to make a fair comparison with RESI. Here, $T_{Hr}$ is the length of the time where Hr is defined and $T_f$ is the length of the period where $f$ is defined. That is, PI and RI are defined for $t > t_1$ and $t > t_0$ respectively as shown in Appendix D, setting $t_0$ and $t_1$ set to January 1983 and January 1987, respectively. The target based on RESI as given by Hr$_{sat}$($S$, $t$) is defined for $t$ larger than $t_0$+ 3 year, i.e., after January 1986, for cutting the first 3 year as noise to consider 3 year or longer period in the ranking of Hr$_{avr}$($S$, $t$) for computing Hr$_{sat}$($S$, $t$). That is, $min$($T$, $t − t_0$) is set to 3 year or longer by giving $T$ larger than 3 year (the difference of $T$ does not affect the performance of the alarm as far as $T$ is larger than 3 year, as shown later in Figure A8). On the other hand, the earthquake activity is not a baseline to compare with RESI but the target to predict. Therefore, high_activity($S$, $t$) is set to 1 if activity($S$, $t$) at time $t$ is larger than its average plus the standard deviation for all the data period and is also the largest in the period of [$t −$ 2 y, $t$]. Thus, months of the significantly larger total magnitude of earthquakes than average, without an earthquake of equal or larger magnitude within two preceding years, are taken.

Then, based on Equations (6)–(8), prec($f$, $g$, $S$, $\Delta t$) and delay($f$, $g$, $S$, $\Delta t$) respectively mean the probability that positive values of high_$f$($S$, $t$) tend to appear earlier and later than high_$g$($S$, $t$) within a time gap of $\Delta t$. Here, $t_e$ represents the ending time of the data i.e., March 2017 here. Thus, prec($f$, $g$, $S$, $\Delta t$) and delay($g$, $f$, $S$, $\Delta t$), defined in Equations (7) and (8) respectively, both imply the possibility to regard $f$ as a precursor of $g$. However, the utility of the two functions are different: prec($f$, activation, $S$, $\Delta t$) suggests to expect the activation of earthquakes in the following period of $\Delta t$ if an alarm defined on $f$ is detected at $t$, whereas delay(activation, $f$, $S$, $\Delta t$) suggests looking back for an alarm in the past $\Delta t$ months, if activation is detected at $t$.

For the larger rate of false positive, the value of function prec($f$, activity, $\Delta t$) in Equation (7) is the lower because the positive value of high_$f$ at time $t$ tends to miss high activities of earthquakes within the period of length $\Delta t$ after $t$. On the other hand, for the larger rate of false negative, the value of delay($f$, activity, $\Delta t$) in Equation (8) is the lower because time $t$ of high earthquake activity tends to miss positive high_$f$ within the period of length $\Delta t$ before $t$. Thus, using Equations (7) and (8) means to evaluate the reliability of RESI for precursor detection based on false positive/negative mentioned at the end of Section 4.

$$\text{unit}(x) = 1 \text{ if } x > 0, \text{ unit}(x) = 0 \text{ if } x = 0 \tag{6}$$

$$\text{prec}(f, g, S, \Delta t) = \frac{\sum_{t=ts}^{te-\Delta t} \text{unit}\Big(\text{high}\_f(S,t) \sum_{\tau=1}^{\Delta t} \text{high}\_g(S, t+\tau)\Big)}{\sum_{t=ts}^{te-\Delta t} \text{unit}\big(high\_f(S,t)\big)} \tag{7}$$

$$\text{delay}(g, f, S, \Delta t) = \frac{\sum_{t=ts+\Delta}^{te} \text{unit}\big(\text{high}\_g(S,t) \sum_{\tau=1}^{\Delta t} \text{high}\_f(S, t-\tau)\big)}{\sum_{t=ts+\Delta}^{te} \text{unit}\big(high\_g(S,t)\big)} \tag{8}$$

In the following evaluation, $g$ and $f$ above are substituted with activity($S$, $t$) and the alarming function which is $\text{Hr}_{\text{sat}}(S, t)$, PI($S$, $t_0$, $t_1$, $t$), or RI($S$, $t_0$, $t_1$, $t$). PI($S$, $t_0$, $t_1$, $t$) represents the estimated risk of earthquakes in region $S$, represented by $\Delta\text{Pi}(t_0, t_1, t)$ in the reference of PI [36,37,40–44]). On the other hand, RI stands for the RI [40,42,45,46]). PI and RI are summarized in Appendix D (referring to [36,37,40,44,45]). An alarming function $f$ is regarded as a precursor of $g$, that peaks earlier than $g$ within the preceding time of $\Delta t$, if prec($f$, $g$, $S$, $\Delta t$) > prec(random, $g$, $S$, $\Delta t$) (Condition A below). $f$ is regarded as a precursor of $g$ for a retrospect from $g$ if delay($g$, $f$, $S$, $\Delta t$) > delay(random, $f$, $S$, $\Delta t$) (Condition B). $\Delta t$ is set to 1 year, 2 year, or 3 year in the following tests. Here, function random($S$, $t$) takes the value of 1 for all $t$, that means to present an alarm at a random time in Condition A. On the other hand, the same function means to assume an imaginary earthquake at a random time in Condition B. Details of the reason for choosing the evaluation method above mentioned in Appendix E referring to the literature [20,75–78]. In summary, the alarming functions are compared as in the following procedure.

The evaluation of the performance (Conditions A and B are two types of successful alarms)
**for each** alarming function in {Hr$_{\text{sat}}$, PI, RI} **do**
　　　$f$ = alarming_function
　　　　　**if** prec($f$, activity, $S$, $\Delta t$) > prec(random, activity, $S$, $\Delta t$): /* Condition A
　　　　　　　A positive value of high_$f$ tends to precede high_ activity in region $S$ within the precedence of $\Delta t$
　　　　　**if** delay(activity, $f$, $S$, $\Delta t$) > delay(random, $f$, $S$, $\Delta t$): /* Condition B
　　　　　　　A positive value of high_activity tends to be after high_$f$ in region $S$ within the delay of $\Delta t$
**end if**
**end for**

The performance of $f$ as an alarming function for earthquake activation is evaluated in Figure 6 based on the values of functions prec and delay with visual representation. Figure 6a–d show the preceding correlation of alarming functions with activity($S$, $t$), for the evaluated period (from 1983 to March 2017). The red dots depict the cells where Conditions A or B turned out to be satisfied, respectively. We find high correspondence of these regions with the 22 active cells (the shadowed cells where the monthly average number of earthquakes of M2.0 or larger is one or larger). That is, RESI substantially outperformed PI and RI for the tested precedence time $\Delta t$ of 1 year and 2 year. For $\Delta t$ of 3 year or longer, the difference in performance is not obvious.

The results of Condition A on the function prec may be improved if we collect earthquakes further in the future. For example, some panels in Figure 5 include peaks of $\text{Hr}_{\text{sat}}(S, t)$ such as (c-3) and (e-3) without any following peaks of earthquake activity. This does not always mean there is no risk, but it may imply the possibility of a forthcoming large earthquake in the region. In Hokkaido, the northernmost island in region (e) but not in (d), we had an M6.7 earthquake in September 2018, within 2 year after (e-3). In this sense, false positive cases should not always be regarded as a failure of precursor detection until we collect data on forthcoming earthquakes.

Regarding Hr($S$, $t$), $T$ used in computing $\text{Hr}_{\text{sat}}$ is set to 28 year because the value of Hr($S$, $t$) tends to increase monotonously for at least 28 year as in Figure A7, so $\text{Hr}_{\text{avr}}(S, t)$ does too. However, as shown in Figure A8, the results of correlation were similar for any $T$ of 3 year or longer for the for $\Delta t$ of 1 year. For the longer precedence, such as 3 year for $\Delta t$, Figure A8 shows that the performance degrades for the shorter $T$. Thus, the superiority of RESI can be expected for $\Delta t$ of 1 year or 2 year, i.e., RESI detects precursors by the finest time resolution.

Indices such as the changing of topic distribution on the dynamic topic models (DTM [79]), not compared with RESI here, may be introduced in the future for the detection of changing points in earthquake history if extended to deal with emerging topics corresponding to emerging latent dynamics which may cause earthquake precursors. However, DTM is not compared with RESI here, because RESI, i.e., Hr($S$, $t$), itself is a function representing the distribution of events to clusters corresponding latent topics obtained in DTM. That is, RESI and DTMs are tools to be combined, rather than compared.

Please note that the values of $\Delta t$ in this comparison have been set to exclude offshoots of past events from the results. That is, this evaluation has been executed setting $\Delta t$ to 12, 24, or 36 months, i.e., 1 year, 2 year, 3 year, for prec($f$, $g$, $S$, $\Delta t$) and delay($g$, $f$, $S$, $\Delta t$). Here the average temporal distance between successive times of non-zero high_activity($S$, $t$), i.e., $\delta t$ where $t + \delta t$ is the nearest time after $t$ where high_activity($S$, $t$) > 0 and high_activity($S$, $t + \delta t$) > 0, is 59.3 months and the standard deviation $\sigma$ is 7.61, averaging for all the 22 active cells mentioned above. Thus, the evaluation comes to be meaningless if $\Delta t$ is set longer than 36 months because ($\delta t - \Delta t$) comes to be less than 3$\sigma$ so it becomes unclear if an obtained alarm is relevant to a future earthquake as a precursor or to a previous earthquake as an offshoot, such as an effect relevant to aftershocks. This is the reason we do not include results for $\Delta t$ longer than 3 year. Although excluding offshoots completely may be impossible, this setting lowers the necessity to consider the overlapping with the impact of previous events.

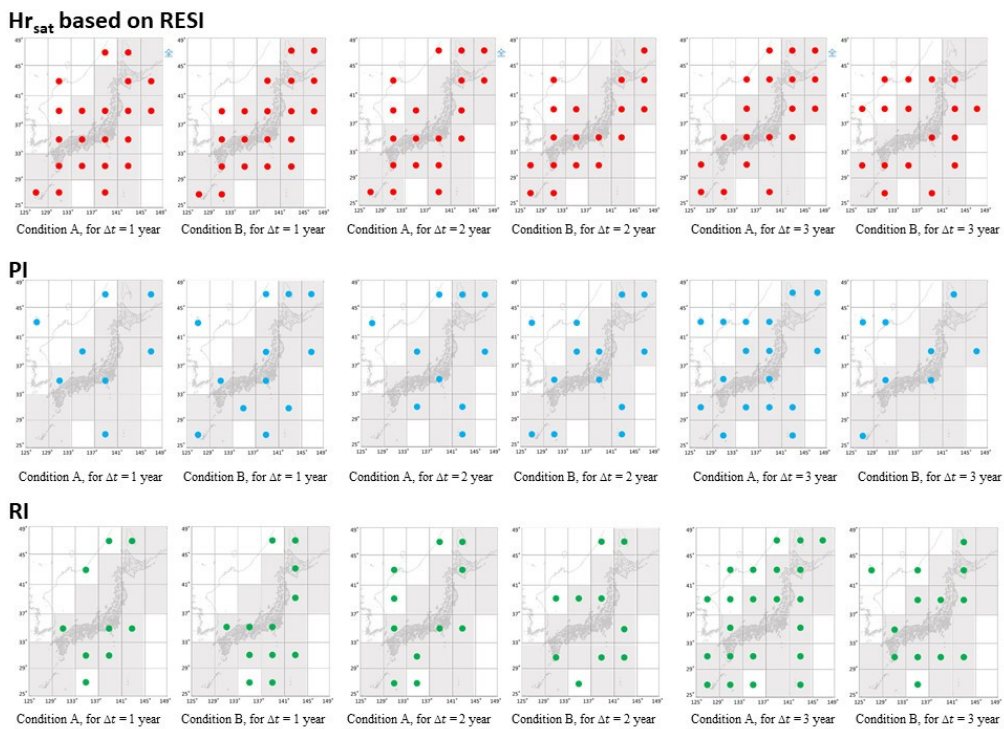

**Figure 6.** The performance of the compared alarming functions. The three alarming functions, i.e., Hr$_{sat}$($S$, $t$), PI($S$, $t$), and RI($S$, $t$), are compared setting $\Delta t$ to 1 year, 2 year, and 3 year. The red dots in a cell mean Condition A stands in the left panel, and Condition B in the right, for each alarming function for each value of $\Delta t$. The shadowed cells show regions where the monthly average number of earthquakes of M2.0 or larger was one or larger. Source of the background map: Japan Meteorological Agency website (http://www.data.jma.go.jp/svd/eqev/data/intens-st/).

## 6. Conclusions

Based on the model of restructuring clusters of quaking meshes, we developed a method to detect precursors of earthquake activation in region $S$ based on an increase in RESI. This means that the average contribution per earthquake in the target region to the diversity of clusters in the whole map was observed. A precursor alarm is detected when an increase in RESI is followed by its saturation or perturbation after the saturation. As a result, RESI substantially outperformed PI and RI for the tested precedence time $\Delta t$ of 1 year and 2 year. In other words, RESI detects precursors by the finest time resolution. The point of RESI is to consider the effect of the distribution of events inside a region (i.e., a cell) to clusters in the region, whereas PI and RI both deal with the seismicity of the region. Because of this difference, RESI reflects the effect of each event in a region on the diversity of the overall map. This occurs, according to the model in this paper, because of the dynamics of clusters to be separated/combined via the appearance/disappearance of seismic gaps. Although we evaluated RESI for the 2D setting here (i.e., on the positions of epicenters and not of foci) for a fair comparison with baselines, the author will extend the method to 3D by including earthquake depth. As a result; however, the computational load will increase.

In addition, to understand the relationship between RESI and other specifications of entropy from the viewpoint of earthquake dynamics, the author attempts to find a setting for a meaningful constructive combination with them. For example, a change in the value of RESI may not only mean one of the transitions of (a), (b), (c), (d), or (e) in Figure 1, but may imply some mixture of them. To capture generalized dynamics in the future work, we may apply and/or develop a theory by extending the stress accumulation model [48] and/or applying the percolation theory [80]. Furthermore, probabilistic models for earthquakes such as ETAS could also be combined. These models may be used to remove the effects of aftershocks from the obtained precursor signals.

*Data and Their Availability*

The target data are taken from the earthquake catalog (Figure A4) provided by the JMA, including the time, latitude, longitude, magnitude, etc. These data are open access at (http://www.data.jma.go.jp/svd/eqev/data/bulletin/hypo.html), the data site of JMA. This is the Japanese version of https://www.data.jma.go.jp/svd/eqev/data/bulletin/shindo_e.html, except that 2017 data (to March) is added. In addition, the location of seismographs is obtained by JMA http://www.data.jma.go.jp/svd/eqev/data/intens-st/, as visualized in Figures 2.

We obtained earthquakes in the data from 1983 until March 2017 of magnitude $M_\theta$ or larger, as in Table A1, setting the cutoff $M_\theta$ to 2.0, smaller than in the literature where 3.0 is used [35,38,81]. The reason for taking a smaller cutoff is because the small earthquakes including aftershocks play an essential role as mentioned in Appendix B, in addition to the reliance of RESI on the distribution of earthquakes of which small ones occupy a large portion. In the sense of the magnitude of completeness ($Mc$), the data of JMA has been taken by high reliability for M2.0 and larger earthquakes. Here, the smallest value in the reliable range i.e., M2.0 or larger has been taken as $M_\theta$.

The values of magnitude and focal positions have been collected in a different regulation before and after 1983 (http://www.data.jma.go.jp/svd/eqev/data/bulletin/data/hypo/relocate.html referring to [82], in Japanese), so we have not used earthquake data before 1983. Although the clustering of quaking meshes can be easily extended to 3-dimensional data, two dimensions i.e., latitude and longitude, are addressed for a fair comparison with baseline methods in Section 5. Other data, that are created on the way for evaluating the values of frequency of earthquakes, RESI, PI, and RI are attached respectively as Supplementary Data S1, S2, S3, and S4.

**Supplementary Materials:** The following are available online at www.mdpi.com/xxx/s1, Data S1, S2, S3, and S4.

**Funding:** This work was funded by JST CREST Grant No. JPMJCR1304, JSPS KAKENHI JP16H01836, and JP16K12428.

**Acknowledgments:** The author appreciates the reviewers for the very constructive comments for improving this paper. Also, the author would like to thank Editage (www.editage.jp) for English language editing.

**Conflicts of Interest:** The author declares no conflict of interest.

## Appendix A. The Convenience of RESI for Dealing with a LARGE-scale Land Crust (Linked from Section 3.1)

**Proposition 1**. *Hr(S⁺, t) is equal to the average of Hr(S, t) for all subregions S of S⁺.*

**Proof.** Given subregions $S_i$ of $S^+$ for $i$ = 1, 2, … $N_S$, where $N_S$ is the number of subregions of $S^+$. The entropy of region $S^+$, which reflects the distribution of earthquakes to clusters of foci in $S^+$, is given as $H$ in Equation (A1). Here, $|S|$ represents the number of clusters in region $S$. $C_{ij}$ represents the $j$-th cluster in sub-region $S_i$. Here, for simplicity, we assume each cluster is properly included in some sub-region, i.e., not divided by sub-region boundaries. Hereafter, "$i \in (1, …, N_s), j \in (1, …, |S|)$" is shortened to "$i, j$".

$$H(S^+) = -\sum_{i,j} p(C_{ij}|S^+) \log p(C_{ij}|S^+)$$
$$= -\sum_{i,j} p(C_{ij})/p(S^+)\{\log p(C_{ij}) - \log p(S^+)\} \tag{A1}$$

Equation (A2) is obtained from Equation (A1) using the equality of $p(C_{ij})$ to $p(C_{ij}|S^+)p(S^+)$ because $C_{ij}$ is a proper part of $S^+$, and $p(S^+)$ to $\sum_{i,j} p(C_{ij})$.

$$p(S^+)H(S^+) = -\sum_{i,j} p(C_{ij}) \log p(C_{ij}) + p(S^+) \log p(S^+). \tag{A2}$$

On the other hand, entropy $H(S_i)$ for region $S_i$ is given as

$$H(S_i) = -\sum_{j} p(C_{ij}|S_i) \log p(C_{ij}|S_i) = -\sum_{j} p(C_{ij})/p(S_i)\{\log p(C_{ij}) - \log p(S_i)\} \tag{A3}$$

Here, the equality of $p(C_{ij})$ to $p(C_{ij}|S_i)p(S_i)$ is used because $C_{ij}$ is a proper part of $S_i$, and $p(S_i)$ to $\sum_{j} p(C_{ij})$. As a result, Equation (A4) is obtained from Equations (A2) and (A3).

$$\sum_{i} p(S_i)\{H(S_i) - \log p(S_i)\} = -\sum_{i,j} p(C_{ij}) \log p(C_{ij}) = p(S^+)\{H(S^+) - \log p(S^+)\}. \tag{A4}$$

Thus, we can say that the RESI of $S^+$, represented by Hr($S^+$) and defined as $\{H(S^+) - \log p(S^+)\}$, is equal to the average of Hr($S_i$) for all subregions $S_i$ of $S$ for $i$ = 1, 2, … $N_S$, weighted by $p(S_i)$. Time $t$ has been just cut above from equations for simplicity, so $H(S_i) - \log p(S_i)$ can be replaced with Hr($S_i$, $t$) according to Equation (2), and $H(S^+) - \log p(S^+)$ with Hr($S^+$, $t$). Then, by substituting each $S_i$ for all $i$ with all $S$ the union of whose cover $S^+$, Equation (A5) is derived.

$$\mathrm{Hr}(S^+, t) = \sum_{S \in S^+} p(S, t)\mathrm{Hr}(S, t)/p(S^+, t). \tag{A5}$$

□

This proposition also justifies the comparison of Hr($S$, $t$) for region $S$ with Hr($S^+$, $t$) for region $S^+$, the background region including $S$ and its surrounding regions, because Hr($S^+$, $t$) can be obtained as the average of the regional entropy of all regions included in $S^+$ using only linear computations. Hr($S$) is called the regional entropy in this paper, i.e., entropy conditioned by region $S$, because Hr($S$) can be extended for wider regions based on this proposition by more simple computation than the original entropy $H(S)$. Generally, the entropy of a system in statistical physics has been known to be the sum (not the average as in this proposition) of entropies of the sub-systems of the system. This difference comes from our definition of regional entropy in which each earthquake belongs exclusively to one cluster, which belongs exclusively to one region.

## Appendix B. Dealing with Small Earthquakes Including Aftershocks (Linked from Data and Their Availability and Appendix C)

There is a possibility that the increase in the number of earthquakes may be just a transient change or aftershocks of a large earthquake. Studies on earthquake predictions often discount or cut aftershocks, regarding them as noise. However, in this paper, aftershocks should not be cut off for the reason below. Here a cluster means a cluster of foci, approximated by clusters of epicenters.

As shown in Figure A1, suppose an earthquake $EQ_A$ occurred at position $u$ at time $t$, and a group of earthquakes as $EQ_C$ occurred as aftershocks of $EQ_A$ in the region. These aftershocks include position $u + du$ at $t + dt$, where $du$ is a vector such that $|du| < \Delta x$ where $\Delta x$ is the mesh size. Also, suppose each earthquake in $EQ_C$ occurred within distance $\Delta x$ from some others in $EQ_C$ (belonging to the same cluster as $EQ_A$). Then, suppose a new earthquake $EQ_B$ at position $v$ at time $t + 2dt$ occurred, belonging to this same cluster, as in (b) of Figure A1, and that the shortest distance from $v$ to one of the focal points in $EQ_C$ is less than $\Delta x$.

Here, note that $EQ_C$ occurred from a cause common to $EQ_A$ because $EQ_C$ is a sequence of aftershocks of $EQ_A$. In addition, $EQ_A$, $EQ_B$, and $EQ_C$ are considered to form one cluster if we consider the line from $u$ to $v$ as a part of a cluster. If we cut away earthquakes on this line because they are small or because they are aftershocks, we overestimate the regional entropy by regarding this cluster as two separate sub-clusters. This creates a result biased to convex clusters, as in Figure A5c.

For example, in the case of earthquakes in Kumamoto prefecture in Japan (M7.3, 2016), the first shock of M6.5 corresponds to $EQ_A$, followed by many small aftershocks as in $EQ_C$, and by the largest earthquake M7.3 corresponding to $EQ_B$. Furthermore, after a sequence of smaller earthquakes, the M5.3 earthquake occurred in the neighboring prefecture, Oita. Among them, the first large quake $EQ_A$, the following largest $EQ_B$, and the one in Oita would be regarded to be in different clusters of foci of frequent earthquakes, if the sequence of small earthquakes between them (as in $EQ_C$) had been cut away from the data.

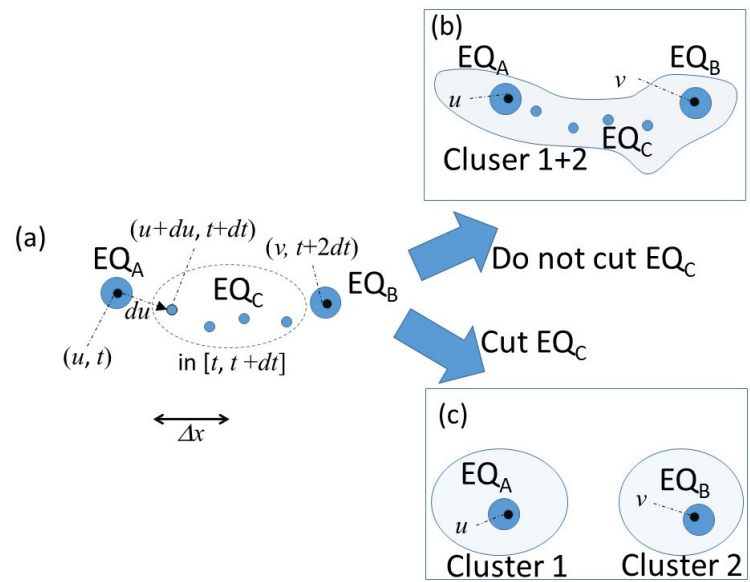

**Figure A1.** The reason aftershocks should not be cut off here. Suppose earthquake $EQ_A$ occurred at position $u$ at time $t$, and a group of earthquakes as $EQ_C$ occurred as aftershocks of $EQ_A$ as in (**a**). Earthquakes in $EQ_C$ occurred close to each other, and a new earthquake $EQ_B$ occurred at position v close to $EQ_C$. $EQ_A$, $EQ_B$, and $EQ_C$ can be regarded as in the same cluster as in (**b**). However, if we cut earthquakes in $EQ_C$, we overestimate the regional entropy, because the cluster gets separated into two as in (**c**). Linked from Appendix B.

### Appendix C. The Clustering Method in Step 1 (Linked from Section 3.2)

The objective of this paper is to detect precursors by computing regional entropy in Step 2. This computation is based on the distribution of earthquakes to the clusters of quaking meshes obtained in Step 1. The method in Step 1 is introduced for the specific purpose of dividing the land surface into clusters separated by gaps composed of meshes where the frequency of earthquakes is low. The comparison with other methods for clustering [69,70] for a more general purpose is out of scope here. However, we can point out **make_clusters** called from Step 1 is guaranteed to run in time and memory of O($n$) for $n$ as the number of all meshes in the map, meaning substantially lower cost than the up-to-date methods for unsupervised learning, such as the unsupervised deep neural network [71].

More importantly, **make_clusters** in Section 3.2 (illustrated in Figure A2) fits the purpose of detecting the changes corresponding to the transitions in Figure 1. That is, the clustering here is required to satisfy an essential condition. A quaking region should be divided into multiple clusters (as in cluster No. 2 and No. 3 in Figure A3a), if and only if they are separated by a one-mesh-wide or a wider gap, to distinguish between the cases of (b) and (c) in Figure A1 (in Appendix B). This means to include such small earthquakes as in EQ$_C$ to the same cluster as the larger ones EQ$_A$ and EQ$_B$ in the vicinity. In other words, the clustering should not suffer from the bias toward convex clusters as $k$-means does (see [69] and its extension to high dimensional data by reducing dimension [70]). The bias toward convex clusters means to obtain only such clusters as in case (c) of Figure A1 even if earthquakes between these clusters exist in the target data as (b) of Figure A1. This difference is abstracted in Figure A3. Approaches such as density- or grid-based clustering take denser regions than their surroundings as clusters, but do not serve the purpose of uniting regions once separated by a boundary but is on the way to be reinforced afterward, if the density on the boundary stay weaker than the centroids of clusters. Although we may avoid bias toward convex clusters by employing spectral clustering [72], its computational complexity is O($n^3$) is not as simple as **make_clusters** where the complexity is O($n$). The pre-setting for using a clustering method (such as of $k$ that should be set initially in $k$-means) is also required in hierarchical clustering where the depth of the hierarchy, meaning how clusters should be divided, is to be determined by a supplementary method or controlled interactively by the user watching the obtained tree.

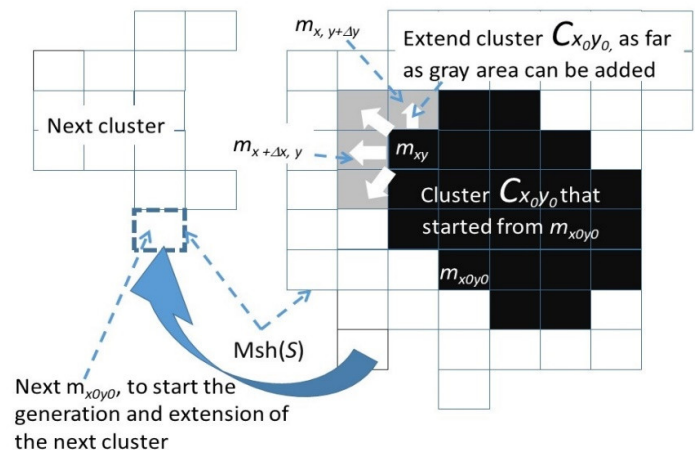

**Figure A2.** Clustering of quaking meshes in Msh($S$) following **make_clusters**. Each cluster ($C_{x0y0}$) grows as a subset of Msh($S$) by tracing neighboring quaking meshes from a seed ($m_{x0y0}$), which is selected randomly from quaking meshes i.e., from Msh($S$), belonging to no cluster generated so far. Therefore, $x_0y_0$, the suffix of the seed mesh, is used as the suffix of cluster $C_{x0y0}$. This growth is halted when no new seed candidate can be found. (Linked from Section 3.2 and Appendix C).

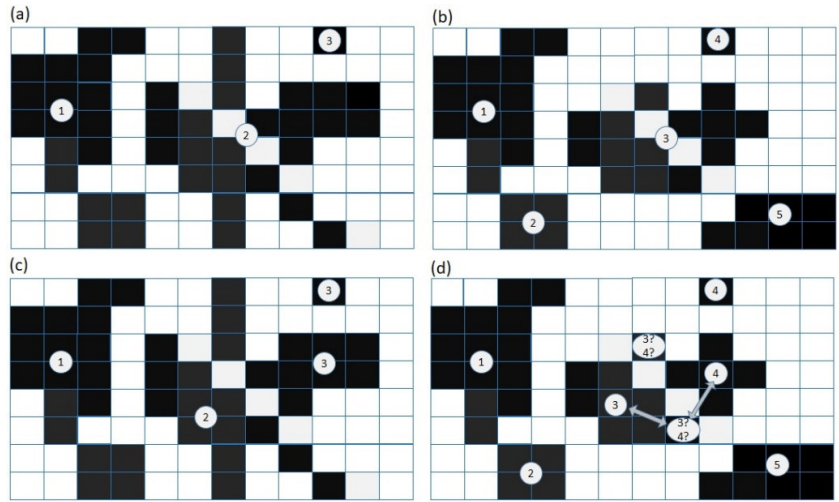

**Figure A3.** Clustering by **make_clusters**($S, t$) in Step 1, compared with $k$-means for various $k$'s. (**a**) and (**b**) show the results of the presented make_clusters for different regions. In (**a**,**b**), each quaking (black) mesh belongs to the same cluster as another quaking mesh, as denoted by the same digit in this figure, as far as they touch via an edge or a vertex. The numbers of clusters as a result of the proposed algorithm from a sequence of earthquakes are 3 for (**a**), 5 for (**b**). On the other hand, (**c**,**d**) show the results of $k$-means for $k$ set to 3 and 5 respectively for regions corresponding to (**a**,**b**). In (**c**), the largest cluster in (**a**) is separated into two, and the mesh isolated in (**a**) (the single-mesh cluster 3) can join the enlarged cluster 3 in the case of (**c**) because the mesh is close to the centroid of this cluster. Furthermore, the cluster 3 in (**a**) is separated into clusters 2 and 3 in (**c**) although they share vertices. Such a result in (**c**) is not acceptable in the proposed method, because the quaking regions separated by a gap of the width of one mesh, such as cluster No./3 in (**c**) should be divided into two clusters, as in clusters 2 and 3 in (**a**). In addition, cluster 3 in (**c**) should be divided into two clusters, such as clusters 2 and 3 in (**a**), for obtaining such a structure as in Figure A1b rather than in Figure A1c. Furthermore, the uncertainty of clusters for some meshes to belong to may occur in the iterative process of $k$-means algorithm, as in (**d**) where two meshes suffer from the equal distance to cluster 3 and 4, which also disturbs analysis. (Linked from Appendix C).

| 10 | 20 | 30 | 40 | 50 | 60 | 70 | 80 | 90 | |
|---|---|---|---|---|---|---|---|---|---|

```
J199510192350135 0  038  281421 165 1303276 203 47    38V      1211 7296NEAR AMAMI-OSHIMA ISLAND  17K
J199510192354154 8  074  280705 196 1302082 315 36    19V      121  7296NEAR AMAMI-OSHIMA ISLAND   3K
J199510192355007 7  020  275286 052 1301313 097 46    18V      121  7296NEAR AMAMI-OSHIMA ISLAND   3K
J199510192354461 5  018  282183 075 1301641 071 275006919V      111  7296NEAR AMAMI-OSHIMA ISLAND   3K
J199510192354460 0  032  334716 330 1365902 438349    28V      171  6208SE OFF KII PENINSULA       5S
J199510192357007 7  046  281592 222 1301606 192 441331624V      111  7296NEAR AMAMI-OSHIMA ISLAND   3K
J199510192358158 1  046  275889 129 1301381 231 51    21V      121  7296NEAR AMAMI-OSHIMA ISLAND   3K
J199510192358565 9  083  280732 231 1302093 393 46    28V      121  7296NEAR AMAMI-OSHIMA ISLAND   4K
J19951020000030 24  029  280326 120 1300648 168 63    24V      121  7296NEAR AMAMI-OSHIMA ISLAND   4K
J199510200001046 2  060  280677 235 1302076 259 32    33V      1212 7296NEAR AMAMI-OSHIMA ISLAND  11K
J19951020000137 23  047  412495 114 1400202 121 3     09V      121  1 160SHIMA PEN REG HOKKAIDO     4K
J199510200002538 1  060  422866 109 1390288 282 14    21V      121  1 18SW OFF HOKKAIDO             8K
J199510200005061 1  062  281062 182 1300210 288 22    20V      121  7296NEAR AMAMI-OSHIMA ISLAND   3K
J199510200005617 0  019  281336 096 1302258 082 23301452 24V      111  7296NEAR AMAMI-OSHIMA ISLAND   3K
J199510200061504 0  043  281669 214 1302405 174 443138227V      111  7296NEAR AMAMI-OSHIMA ISLAND   3K
J199510200074969 8  115  281066 272 1301816 506 31    32V      1713 7296NEAR AMAMI-OSHIMA ISLAND   5S
```

| date and time | | error | latitude | error | longitude | error | depth | error | magnitude | | epicenter by the name of place |

**Figure A4.** The data structure in the earthquake catalog of JMA. Earthquakes of JMA magnitude (the two digits xy above means magnitude x + y/10) 2.0 or larger are dealt with, where magnitude values have been obtained by JMA. The errors follow the corresponding data attribute: for example, the first "error" above is the error of "time and date" shown by three digits representing seconds. Source: Japan Meteorological Agency website (http://www.data.jma.go.jp/svd/eqev/data/bulletin/hypo.html). (linked from **Data and Their Availability**).

**Table A1.** The number of earthquakes of M2.0 or larger in the target. The target area is (25,125)–(45,145) in Figure 2, for each year in the target of the analysis. The total number for all years is 613,136. (linked from **Data and Their Availability**).

| Year | The Num. of Earthquakes | | | | | | |
|------|------|------|--------|------|--------|------|--------|
|      |      | 1991 | 6473   | 2001 | 15,908 | 2011 | 88,562 |
|      |      | 1992 | 8702   | 2002 | 15,038 | 2012 | 35,812 |
| 1983 | 4807 | 1993 | 13,234 | 2003 | 20,222 | 2013 | 24,821 |
| 1984 | 4168 | 1994 | 15,167 | 2004 | 21,301 | 2014 | 21,360 |
| 1985 | 3970 | 1995 | 25,063 | 2005 | 19,838 | 2015 | 19,989 |
| 1986 | 4874 | 1996 | 16,855 | 2006 | 15,436 | 2016 | 33776  |
| 1987 | 5204 | 1997 | 16,212 | 2007 | 17,085 | 2017 | 7364   |
| 1988 | 5448 | 1998 | 16,038 | 2008 | 18,695 |      |        |
| 1989 | 6537 | 1999 | 13,773 | 2009 | 16,015 |      |        |
| 1990 | 6307 | 2000 | 32,423 | 2010 | 16,659 |      |        |

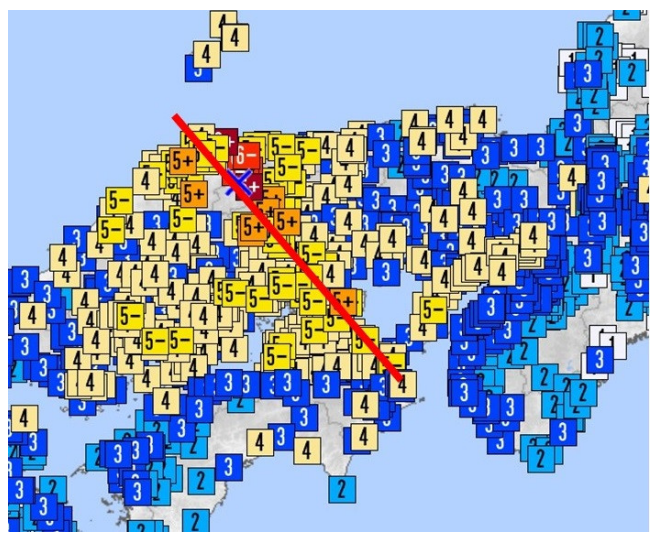

**Figure A5.** The distribution of earthquake intensity. The digits mean the intensity by the earthquake on 6 October 2000. Source: Japan Meteorological Agency website https://www.data.jma.go.jp/svd/eqdb/data/shindo/Event.php?ID=156371. (Linked from Section 4.2).

## Appendix D. The Outlines of Compared Baselines (Linked from Section 5)

In this section, abstracts of PI and RI, compared with the proposed index, are introduced to explain the usage of these methods in the experiments. The details are referred to the original literature shown below.

**Index in Pattern Informatics PI($S_i$, $t_0$, $t_1$, $t$):** PI represents the risk of earthquakes in region $S_i$, represented by $\Delta P_i(S_i, t_0, t_1, t)$ following the literature in PI [36,37,40–44]. Here, as illustrated in Figure A6, [$t_0$, $t_1$] is the reference time range in the past, and [$t$, $t + \Delta t$] is the target period to predict the risk. For each time $t_b$ in [$t_0$, $t$], the average frequency of events above a defined cutoff ($M_\theta$ set to 2.0, for a fair comparison with RESI) from $t_b$ to $t$ is obtained. This average frequency is normalized (i.e., divided by the standard deviation for all regions $S_i$ for all $i$, after subtracting the average for all $S_i$) to obtain $I_i$ ($t_b$, $t$). Then $\Delta I_i(t_b, t_1, t)$ is obtained as $I_i(t_b, t) - I_i(t_b, t_1)$. The average of $\Delta I_i(t_b, t_1, t)$ for all $t_b$ in [$t_0$, $t_1$] is computed to obtain $<\Delta I'_i(t_b, t_1, t)>$. Finally, $<\Delta I_i(t_0, t_1, t)>$ is squared and its difference from the background value (average for all $S_i$ in the given map) is obtained as PI($S_i$, $t_0$, $t_1$, $t$) where $t_0$ and $t_1$ are set to be Jan 1983 and Jan 1987 respectively in this paper. Summarizing this method, the risk in the period [$t$, $t + \Delta t$] in region $S_i$ is estimated as the change in the earthquake risk for the recent period from $t_1$ and $t$.

**Relative intensity RI($S_i$, $t_0$, $t_1$, $t$):** For each time $t$, RI (for details see [40,42,45,46]) here is given by $ni$ ($t_0$, $t_1$, $M_\theta$)/$\sum_{S_j \subset S^U} nj$ ($t_0$, $t_1$, $M_\theta$). Here, $ni$ ($t_0$, $t_1$, $M_\theta$) is the average number of earthquakes in the cells including $S_i$ and the adjacent cells surrounding $S_i$. $M_\theta$ is the cutoff magnitude and $t_0$ and $t_1$ are times before $t$, the target period of earthquake prediction (precisely $[t, t + \Delta t]$ is the target period by setting $\Delta t$ to 1 year or 1 month in this paper). $S^U$ is the set of all the cells in the given map corresponding to the data analyzed. In applying RI in this paper, $t_0$ and $t_1$ have been said to be ideal if $t_1 - t_0$ is set large [45], so here we set $t_0$ as January 1983 and $t_1$ as $t - 1$.

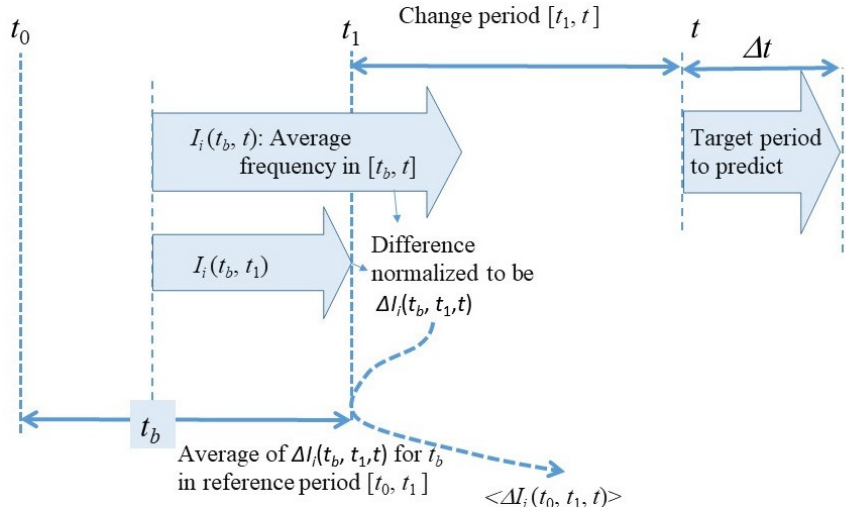

**Figure A6.** Computing $<\Delta l_i (t_0, t_1, t)>$ in obtaining $\mathrm{PI}(S_i, t_0, t_1, t)$. PI is the index in pattern informatics that is compared with $\mathrm{Hr}(S_i, t)$, where $[t_0, t_1]$ is the reference period and $[t, t + \Delta t]$ is the target period of which the risk of earthquakes should be predicted. Note: the symbols used here differ from those in the references. For example, $t$ is given as $t_2$ in the literature such as [36,37,40]. Linked from Appendix D.

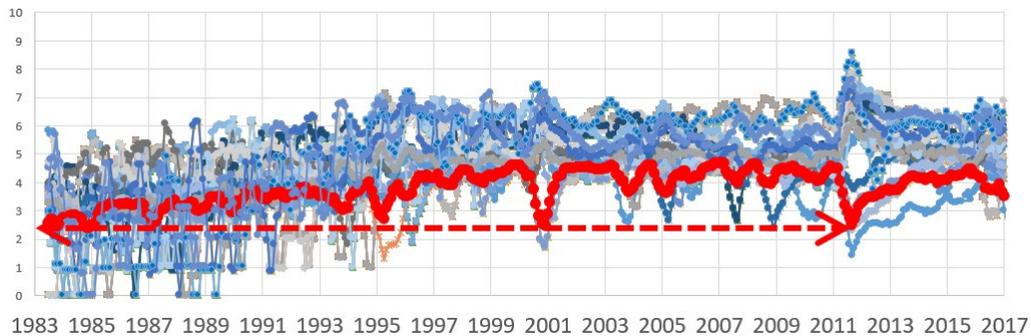

**Figure A7.** The monthly transition of RESI for all regions and all Japan. $\mathrm{Hr}(S, t)$ for all S i.e., 36 regions, corresponding to the cells in Figure 2 shown by the thin blue background curves. The thick red curve is obtained as the average of the 36 regions by Equation (A5) in Appendix A, i.e., the value transition of RESI for the entire map of 36 regions. We find the increasing trend in the thick average curve for 28 year as in the dotted arrow. (Linked from Section 5).

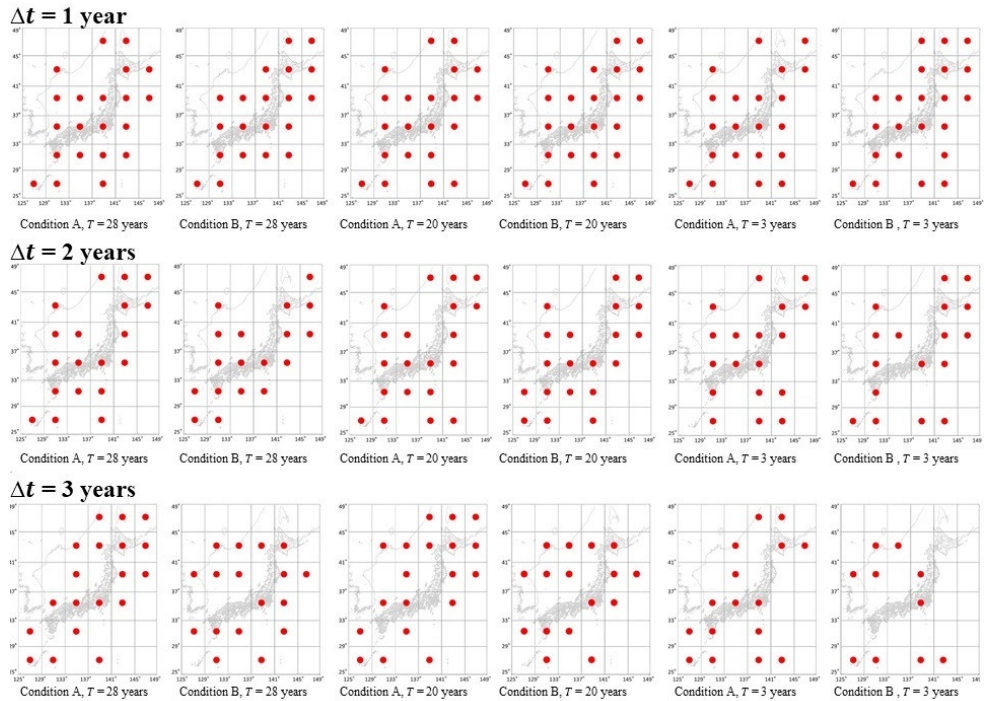

**Figure A8.** The performance of $\mathrm{Hr_{sat}}(S, t)$ for various setting of $\Delta t$ and $T$. The alarming function $\mathrm{Hr_{sat}}(S, t)$, is compared for setting $\Delta t$ to 1 year, 2 year, and 3 year respectively in (**a–c**) here, setting $T$ in computing $\mathrm{Hr_{sat}}$ to 3 year, 20 year, and 28 year. The result did not change substantially when $T$ was set any longer ($T$ is shorter than 35 year, which is the period of the available data). The red dots in a cell mean the same as in Figure 6. Source of the background map: Japan Meteorological Agency website (http://www.data.jma.go.jp/svd/eqev/data/intens-st/) (Linked from Section 5).

## Appendix E. The Method for Evaluating Alarming Functions (Linked from Section 5)

This section is presented to explain the reason for choosing the evaluation criteria of Equations (7) and (8) in the comparison of RESI with baselines. Here, let us show some possible alternative methods for evaluation and discuss the advantages and disadvantages.

An alternative method for evaluating the correlation between an alarming function and the activation of earthquakes is to take the point-to-point similarity of two sequences ($f$: alarming function, $g$: earthquake activity), where the times of the corresponding points in the two series may not always correspond exactly. This evaluation may be executed by using Dynamic Time Warping (DTM [75]), for which the availability as an existing tool is an advantage. However, in this paper, the aim is to evaluate not only the similarity but also the correlation of the obtained precursory alarms, within the precedence of a given length of time ($\Delta t$), with the activation of earthquakes. Therefore, Equation (7) or (8) is used for evaluating the correlation of two sequences within the time gap of $\Delta t$.

Another prevalent evaluation measure is the correlation of the alarms with the real activations with tolerant delay, as is performed in the literature [20]. For example, the precedence of obtained alarms before the target events have been measured with the rate of error (false positive/negative) [76]. However, the purpose of the evaluation in this paper is to investigate the correlation of function $f$ representing the alarm of precursor, within a given time of precedence, with $g$ representing the real activation of earthquakes. To fit this purpose, the criteria in Equations (7) and (8) are used to evaluate the correlation with a tolerant delay $\Delta t$ or with a precedence time $\Delta t$, for a given $\Delta t$ (1 year or 2 year in the evaluation in this paper).

The Area Under Curve (AUC) of Receiver Operating Characteristic (ROC) has also been employed as a measure of an alarm's correspondence to the target event. AUC integrates the evaluations for various values of threshold $\theta_f$, on which a two-dimensional curve is drawn where $X$ takes the conditional probability $p(f > \theta_f \wedge not \text{ high\_ } g \mid not \text{ high\_ } g)$ and $Y$ takes $p(f > \theta_f \wedge \text{high\_ } g \mid \text{high\_} g)$. The width of the area under this curve is AUC. AUC has been quite well used for the evaluation of alarms for diagnosis and detection of signs [20,61,77,78]. However, in this specific case to evaluate RESI, AUC is not applied because of the feature of $\text{Hr}_{\text{sat}}$. That is, as in the main text, the times for the alarming is taken on the combined conditions that is

$\text{rank}_{\tau \text{ in } [t\text{-min}(T,\ t-t0),\ t]}\ \text{Hr}_{\text{avr}}(S, \tau = t) \leq \gamma \min(T, t - t_0)$ **and**

$(\text{stdev}_{\tau \text{ in } [t-dt,\ t]}\ \text{Hr}(S, \tau) < \theta_{\text{std}}$ **or** $\text{stdev}_{\tau \text{ in } [t-dt/2,\ t]}\ \text{Hr}(S, \tau) > 2\ \text{stdev}_{\tau \text{ in } [t-dt,\ t-dt/2]}\ \text{Hr}(S, \tau))$

instead of applying a threshold assuming the monotonic dependence of alarms on the value of Hr. This is the difference from the choice of alarms on the condition $f > \theta_f$ above in AUC. In fact, at the times chosen for alarms in Figure 4 (e.g., (b-2) and (c-2)) or Figure 5 (e.g., (a-2) and (e-2)), Hr is not always larger than times not chosen. Although we can change the number of chosen alarm times ($t$ here) by changing the values $\gamma$ and $\theta_{\text{std}}$ (and the coefficient for stdev from 2 to other values) above, the curve cannot be drawn so simply as the curve drawn usually for AUC. For this reason, $\gamma$ and $\theta_{\text{std}}$ are set to constant values and the functions prec() and delay() are employed that correspond to evaluating based on false positive and false negative (as discussed in Section 5) that is a known merit of employing AUC for evaluation.